\documentclass[aps, twocolumn, superscriptaddress]{revtex4-1}

\usepackage{graphicx}
\usepackage{amsmath}
\usepackage{dsfont}
\usepackage{amssymb}
\usepackage{physics}
\usepackage{hyperref}
\usepackage{ulem}
\hypersetup{colorlinks=true,
	    final=true,
	    linkcolor=blue,
	    citecolor=blue,
	    filecolor=blue,
	    urlcolor=blue,}

\newcommand{\beq}{\begin{equation}}
\newcommand{\eeq}{\end{equation}}

\newcommand{\mathp}{\mathbf{p}}
\newcommand{\mathv}{\mathbf{v}}

\newcommand{\mathE}{\mathbf{E}}
\newcommand{\mathH}{\mathbf{H}}

\begin{document}

\title{A review: The Gor'kov-Teitel'baum thermal activation model for cuprates}
\author{Navinder Singh}
\email{navinder.phy@gmail.com; +919662680605}
\affiliation{Theoretical Physics Division, Physical Research Laboratory, Ahmedabad, India. PIN: 380009.}

\begin{abstract}
While closing their famous paper entitled "Pseudogap: friend or foe of high-Tc?" Norman, Pines, and Kallin underlined that before we have a microscopic theory, we must have a consistent phenomenology. This was in 2005.  As it turns out in 2006 a phenomenological theory of the pseudogap state was proposed by Gor'kov and Teitel'baum. This originated from their careful analysis of the Hall effect data, and it has been very successful model as numerous investigations over the years has shown. In this mini-review the essence of the idea of Gor'kov and Teitel'baum is presented. The pseudogap obtained by them from the Hall effect data agrees very well with that obtained from the ARPES data. This famous Gor'kov-Teitel'baum Thermal Activation model (in short GTTA model) not only presents a consistent phenomenology of the pseudogap state but also it rationalizes the Hall angle data and it presents a strong case against the famous "two-relaxation times" idea of Anderson and collaborators. 
\end{abstract}

\maketitle

\section{Introduction}
The Hall coefficient:
\beq
R_H = \frac{1}{n e}
\eeq
 is roughly temperature independent for simple metals. But for cuprates it exhibits marked temperature dependence (figure 1): decreasing with increasing temperature in the underdoping and optimal doping regimes. The question that has been asked is: how do we understand the temperature dependence of the Hall coefficient for cuprates? All the parameters that appear in equation (1) are apparently constant. Then how does the temperature dependence originate? But before that, one can argue that the applicability of equation (1) to cuprates is itself doubtful and is to be questioned. Within the free electron model and Drude theory or using Boltzmann equation for parabolic band, one can obtain equation (1)\cite{ash}. As argued by Gor'kov and Teitel'baum\cite{gt1} and as shown by\cite{kho} the expression (equation (1)) goes through even for interacting electrons with isotropic energy spectrum. It also goes through in the limit of strong magnetic fields\cite{lif}. Setting aside the theoretical deductions,  there is {\it a posterior} justification for the validity of equation (1)  at least in the underdoped LSCO. The number of carriers doped in $CuO_2$ planes (estimated, for example, from $Sr$ concentration in $La_{2-x} Sr_x CuO_4$) matches with that obtained from equation (1) in the underdoped regime. This was shown by Gor'kov and Teitel'baum in 2006\cite{gt2} (refer to their figure 1). Thus equation (1) can be trusted for underdoped LSCO, and at other dopings it will yield only an {\it effective} number of carriers. 
 
 \begin{figure}[h!]
    \centering
    \includegraphics[width=0.8\columnwidth]{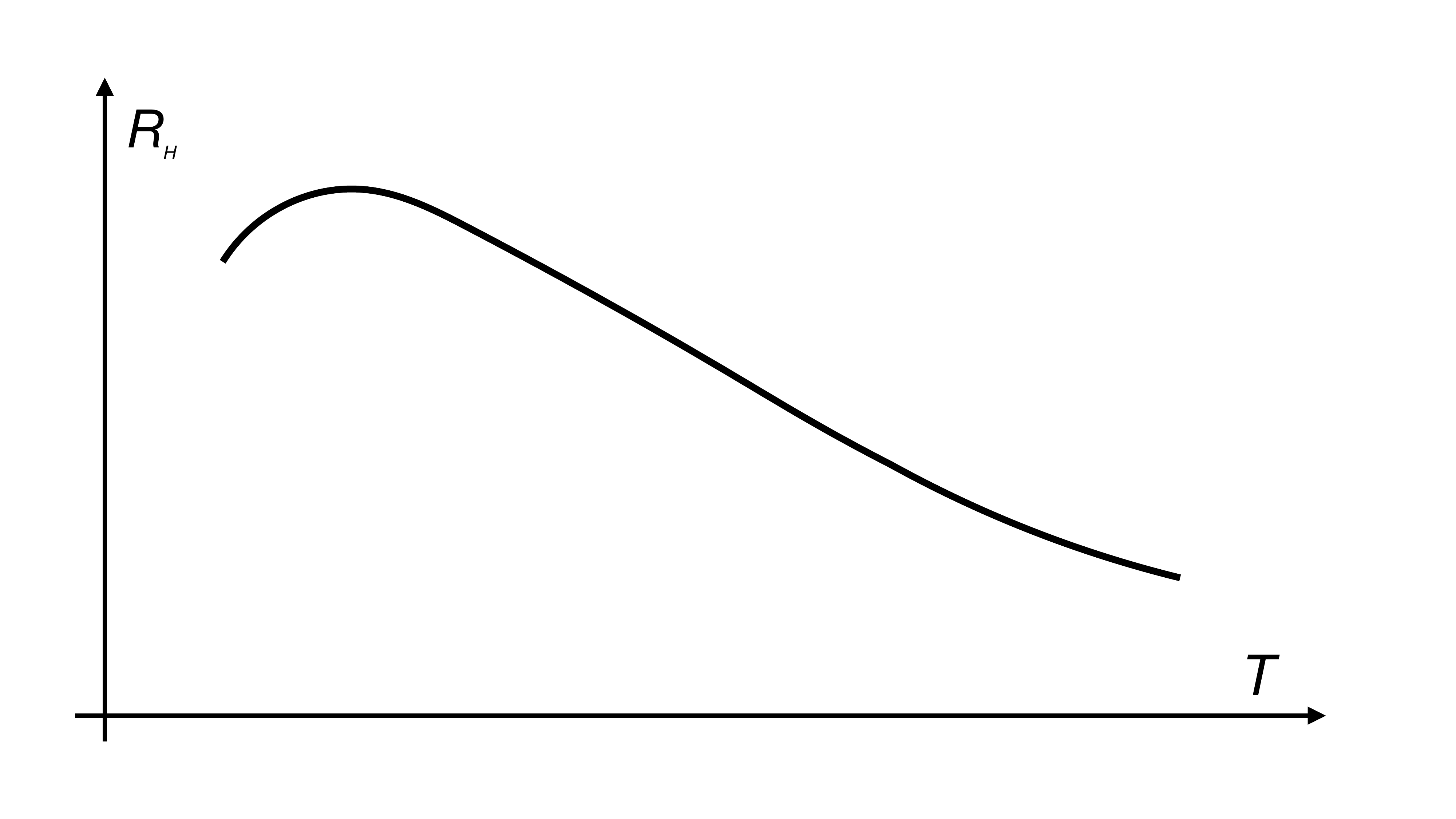}
    \caption{Schematic behaviour of Hall coefficient with temperature in underdoped cuprates.}
    \label{f1}
\end{figure}
 
With this understanding in mind and within the leeway of equation (1), there are two main approaches using which one can rationalize the temperature dependence of the measured Hall coefficient in cuprates: 

\begin{enumerate}
\item Keep $n$ fixed but introduce two different relaxation times: one for transport perpendicular to the Fermi surface and other parallel to it, and attach different temperature dependences to them (an approach followed by Phil Anderson and collaborators)\cite{ander1, chien}.
\item Attach temperature dependence to $n$ in equation (1) (an approach followed by Gor'kov and Teitel'baum\cite{gt2}, refer also to\cite{ono,fot0}).
\end{enumerate}

In what follows, we discuss these approaches one by one. It turns out that the approach adopted by Gor'kov and Teitel'baum  (the GTTA model) addresses the Hall data in a more consistent way and it also offers a phenomenological model of the pseudogap state.  The pseudogap energy scale obtained using GTTA model agrees very well with that obtained from ARPES experiments. As we will see in section (IV) Gor'kov-Teitel'baum approach offers answers to other connected mysteries of the PG sate. In section (V), some other observations by Gor'kov and Teitel'baum are briefed. In section (VI) some approaches that do not fall under the leeway of equation (1) are mentioned. We end the manuscript with a summary in section (VII).

\section{Anderson's way}
Let us consider the Drude theory version of the Anderson's idea (for Hidden Fermi Liquid Theory version of it, refer to\cite{nav}). The equation of motion for an electron under the action of electric and magnetic fields is given by
\beq
\frac{d\mathp}{dt} = -e (\mathE +\frac{1}{c} \mathv\times \mathH) -\frac{\mathp}{\tau}.
\eeq
Here $\tau$ is the average momentum relaxation time. In steady state,
\beq
\frac{\mathp}{\tau} = -e (\mathE +\frac{1}{c} \mathv\times \mathH).
\eeq
Instead of the one relaxation time $\tau$, Anderson introduced two different relaxation times (called "two-relaxation times" scenario) one for scattering due to electric field ($\tau$) and another for scattering due to magnetic filed ($\tau_H$). This can be written as

\begin{eqnarray}
\frac{\mathp_E}{\tau} &=& - e \mathE\nonumber\\
\frac{\mathp_H}{\tau_H} &=& -\frac{e}{c} \mathv\times H\nonumber\\
\mathp = \mathp_E + \mathp_H &=& -e\tau \mathE -\frac{e}{c}\tau_H \mathv\times \mathH
\end{eqnarray}
In the simple Drude theory (when one sets $\tau =\tau_H$) the above equation reduces to equation (3). Writing equation (4) in component wise, and introducing current densities ($J_x = -nev_x,~~J_y = -nev_y$), one obtains:

\begin{eqnarray}
J_x &=& \sigma_0 E_x -\omega_c \tau_H J_y\nonumber\\
J_y &=& \sigma_0 E_y + \omega_c\tau_H J_x.
\end{eqnarray}

Here $\sigma_0 =\frac{ne^2\tau}{m}$ and $\omega_c = \frac{eH}{mc}$ are DC conductivity and cyclotron frequency, respectively. Implementing the open circuit condition (as applicable in an experimental setting): $J_y =0$, one obtains the Hall coefficient:

\beq
R_H = \frac{E_y}{J_x H} = -\frac{1}{nec}\frac{\tau_H}{\tau}.
\eeq

This expression differs from that in equation (1). It has an extra factor of $\frac{\tau_H}{\tau}$. Anderson argued that for charge transport perpendicular to the Fermi surface the scattering rate is given by $\frac{1}{\tau}\propto T$ and for charge transport parallel to the Fermi surface (due to magnetic field as magnetic field only changes the direction of momentum not its magnitude) a different relaxation rate is introduced: $\frac{1}{\tau_H} \propto T^2$. Implementing these assumptions in the above equation for  the Hall coefficient it is clear  that $R_H\propto \frac{1}{T}$. This qualitatively reproduces the data (compare for rough comparison with figure 1)\cite{fot1}. However, a better agreement is obtained by Gor'kov and Teitel'baum using their thermal activation model as discussed in the next section.

\section{Gor'kov-Teitel'baum's way}

\begin{figure}[h!]
    \centering
    \includegraphics[width=0.7\columnwidth]{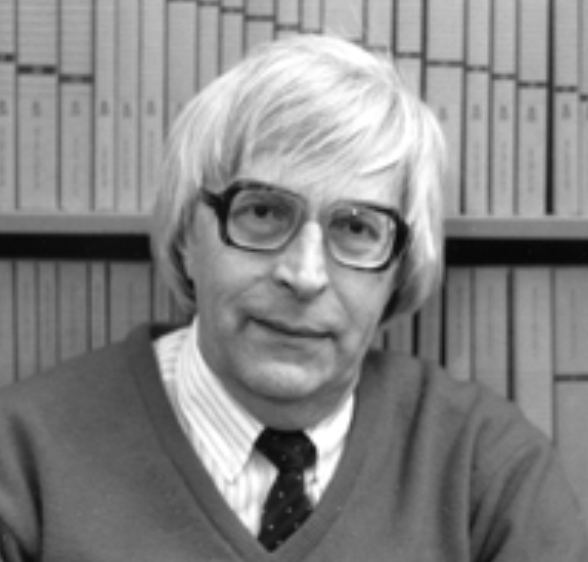}
    \caption{Through this article author pays his tribute to Lev P. Gor'kov (14th June 1929 -- 28th Dec 2016) who with his collaborators did significant and pivotal work regarding the pseudogap state of cuprates and related issues. He has many other important contributions including establishing the microscopic foundations of the Ginzburg-Landau theory.  Image courtesy: Niels Bohr library and archives.}
    \label{f1}
\end{figure}

Gor'kov and Teitel'baum instead of assuming $n$ to be a fixed quantity in equation (1) assumed that it is a doping and temperature dependent function. The so called the Gorkov-Teitelbaum Thermal Activation (GTTA) model is motivated in the following way\cite{gt2}. In their original paper the Hall effect data of \(La_{2-x}Sr_{x}CuO_{4}\) \cite{gt2} was rationalized using this phenomenological model:

\begin{equation}
    R_{H}=\frac{1}{n_{Hall}(x,T) e}
    \label{GTTA}
\end{equation}
\begin{equation}
    n_{Hall}(x,T)=n_{0}(x)+n_{1}(x)e^{-\Delta(x)/T}.
    \label{GTTA}
\end{equation}

Here, the first term, \(n_{0}(x)\), is a temperature independent term which corresponds to externally doped holes. The second term, $n_{1}(x)e^{-\Delta(x)/T} $, is a temperature dependent term which is to be interpreted as thermally activated carriers above a doping dependent energy gap $\Delta(x)$. This is the pivotal term which brings temperature dependence to the Hall coefficient. The first term, \(n_{0}(x)\) varies roughly linearly with $x$ in the  low doping regime (up to \(x\simeq0.07\)) in LSCO (refer to the schematic diagram in figure 3 or refer to original diagram (figure 1) in\cite{gt2}). For higher dopings, \(x>0.07\), \(n_{0}(x)\)  is no more linear and increases super-linearly with $x$. This seems like "opening of a quantum jam" in which, initially, the number of carriers available for conduction are equal to the externally doped carriers and then more and more carriers become available for conduction when doping is increased further (this is at a fixed temperature). In the second term, the pre-factor \(n_{1}(x)\simeq2.8\) is roughly a constant for a wider doping concentrations (\(x<0.19\)) and then it drops very abruptly for (\(x>0.19\)). This behaviour of  $n_1$ in the thermal activation term (second term in the GTTA model) points towards a very important aspect. The thermal activation term appears to significantly weaken above the critical doping $x>0.19$. This means that energy gap $\Delta(x)$ itself appears to close at around $x=0.19$. 

 \begin{figure}[h!]
    \centering
    \includegraphics[width=1.0\columnwidth]{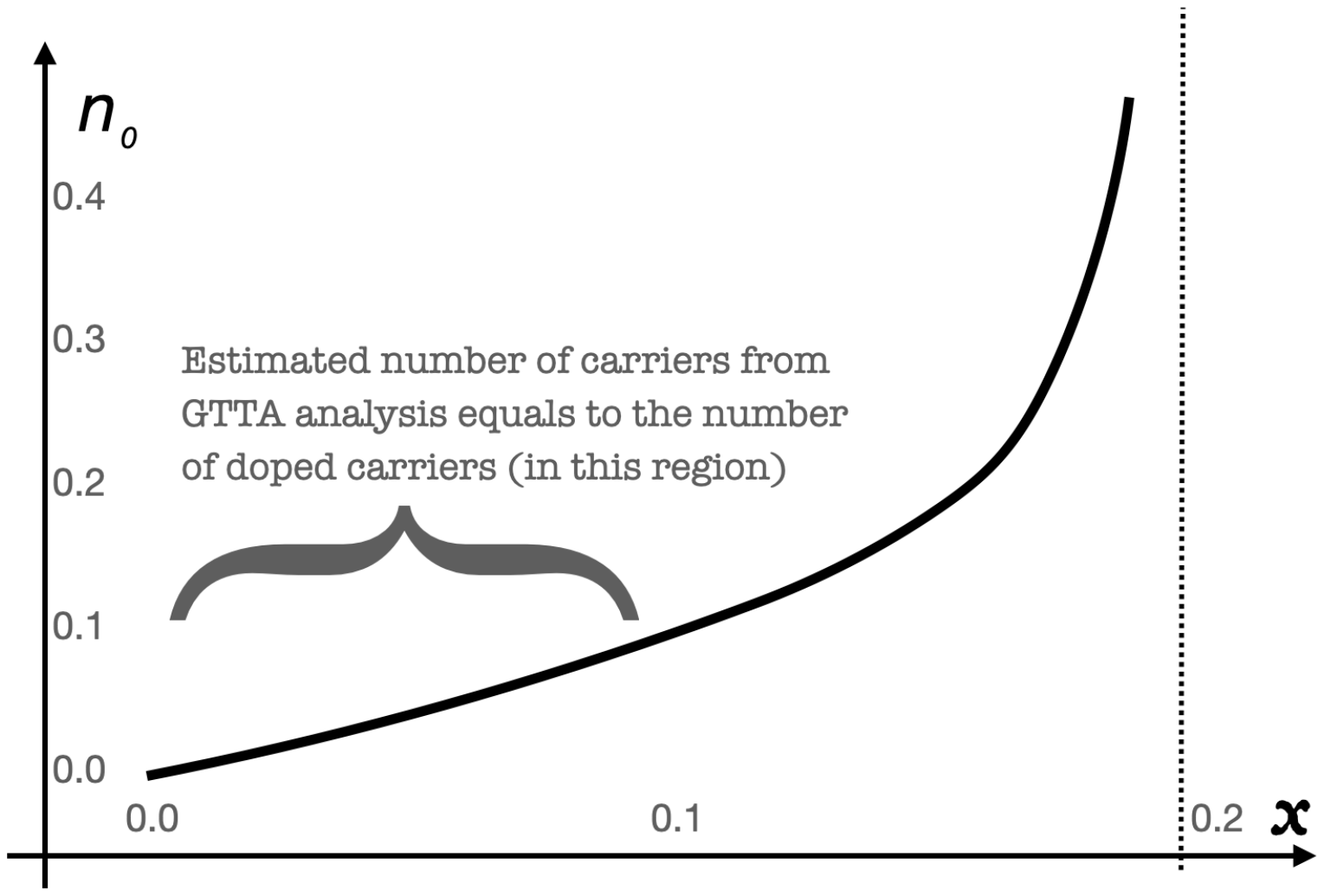}
    \caption{Schematic behaviour of $n_0(x)$.}
    \label{f2}
\end{figure}

The behaviour of $n_1$ is also reconcilable with the carrier density change observed around $p^*$\cite{}. When doping is increased from $p<p^*$ to $p>p^*$ (that is through the quantum critical point at $p^*$), carrier density obtained from Hall coefficient also "jumps" from $p$ to $1+p$. This is as if "some tied down carriers are liberated and set free" when $p$ is increased through $p^*$. And they do not face an energy barrier for thermal activation. What was holding back or tying down these carriers? It is the quasi-localization of carriers on copper sites and the resulting pseudogap due to quasi-localization\cite{pro,nav2}?

 The $\Delta(x)$ is referred as the activation energy required to excite carriers, and it corresponds to the energy difference between the "Fermi arc" (along the nodal directions in the Brillouin  zone on which conducting carriers "lives") and the band bottom along the anti-nodal directions\cite{gt2}. It turns out that the extracted values of $\Delta(x)$ at various doping concentrations agrees very well with that obtained from a completely different experiment (that is ARPES). The activation gap $\Delta(x)$ decreases roughly linearly with increasing \(x\) upto  \(x\simeq 0.20\). Therefore, the \(\Delta(x)\) can be identified with the pseudogap energy scale (refer to figures 2 and 3 in\cite{gt2}).

 \section{The enigma of Hall angle $\cot\Theta_H$}
 
 The Hall angle is defined as $\tan\Theta_H = \frac{E_y}{E_x}$. Here $E_x$ is the applied electric field along the x-axis and $E_y$ is the induced Hall field along the y-direction. Under the open circuit condition ($J_y =0$), the first equation from the equation array (5) leads to the equation $J_x = \sigma_0 E_x$, and the second equation of that array leads to $\sigma_0 E_y = -\omega_c\tau_H J_x$. Combining these two resulting equations, the ratio $\frac{E_y}{E_x}$ is equal to $\omega_c \tau_H$ in magnitude, or
 
 \beq
 \cot\Theta_H =\frac{1}{\omega_c \tau_H} \propto T^2.
 \eeq

As $\frac{1}{\tau_H} \propto T^2$ in Anderson's theory. Experimental verification of it in the early 1990s presented a strong case in favor of Anderson's ideas\cite{chien}. However, alternative approaches were also presented (refer to section (VI)).

Although Gor'kov and Teitel'baum did not solve this problem but hinted at a very novel way of looking at this problem\cite{gt1}. They argued that the number density of carriers (equation (8)) is both doping and temperature dependent. And attaching temperature dependence {\it only} to the transport relaxation rate in
\beq
\rho_{ab} =\frac{m^*}{n e^2}\frac{1}{\tau(T)}
\eeq
may not be sufficient. The number density of carriers sitting in the above formula is temperature dependent. Thermal activation of carriers in the underdoped cuprates can not be neglected. They suggested: 
\beq
\rho_{ab}(T) = \frac{m^*}{n_{Hall}(x,T) e^2}\frac{1}{\tau(x,T)}.
\eeq
And plotted
\beq
\frac{m^*}{e^2}\frac{1}{\tau(x,T)} = \rho_{ab}(T) n_{Hall}(x,T).
\eeq
On purely experimental basis, temperature dependence of $\frac{1}{\tau}$ can be determined, if we experimentally know $\rho_{ab}$ and if we deduce $n_{Hall}(x,T)$ from GTTA model from experimentally known Hall data (and assume $m^*$ to be temperature independent). It is a very surprising fact that they observe
\beq
\frac{m^*}{e^2}\frac{1}{\tau(x,T)} \sim T^2
\eeq
to a good approximation in the case of LSCO  (refer to their figures (1) and (2) \cite{gt1}). Thus there seems to be just one relaxation time (at least in the case of underdoped LSCO).  This relaxation rate which is quadratic in temperature is observed through $\cot\Theta_H$. Do we still need two relaxation times? The result presented by Gor'kov and Teitel'baum is directly coming from the experiments (GTTA model has its full validity verified in so many ways\cite{gt3}). When $n_{Hall}$ from magnetic data is combined with electronic transport data $\rho_{ab}$ we get a transport relaxation rate which is quadratic in $T$ not linear in $T$. What happens at the optimal doping? Is it legitimate to speak about two relaxation times there? This tricky state of affairs is further taken up by the authors and results will be discussed soon\cite{nav3}.

 \section{Some other observations of Gor'kov and Teitel'baum}

 \subsection{Their views regarding the pseudogap state}
 
On the right hand side of equation (8) there are two terms: one is temperature independent term and at low doping it reflects externally doped holes, and other term is the thermal activation term (temperature dependent term). The crossover between strange metal regime to the pseudogapped regime, according to Gor'kov and Teitel'baum, is determined by a balance between these two terms: 
 
 \beq
 n_0(x) \simeq n_1 e^{-\Delta(x)/T^*}
 \eeq
 
 This sets a doping dependent temperature scale:
 
 \beq
 T^* \simeq \frac{\Delta(x)}{\ln(n_1/n_0(x))}.
 \eeq
 
Here $T^*$ is measured in energy units and is the crossover temperature.  The doping dependent gap $\Delta(x)$ is interpreted as an activation energy (energy difference between the Fermi surface "arcs" (along the nodal directions) and band bottom (along the anti-nodal direction))\cite{gt2}. A simple formula (valid in the underdoped regime) can be obtained by setting $n_0(x) \simeq x$ and $n_1\simeq 2.8$ which is roughly given as

\beq
T^*(x) \sim -\frac{\Delta(x)}{\ln(x)}.
\eeq
 
This high energy scale is the same where static magnetic susceptibility exhibits a maximum. There is a lower energy pseudogap where nematicity appears (refer to\cite{nav2} and references therein). According to Gor'kov and Teitel'baum, PG energy scale is set by a balance between thermal activation term and temperature independent term in equation (8), and the doping dependent thermal activation energy $\Delta(x)$ leads to a temperature scale $T^*$ (equation (16)) -- called the PG scale\cite{gt2}.

 \subsection{Their views regarding the electron pocket at the $\Gamma$ point}
 
 In quantum oscillation experiments (QO) an electron pocket at $\Gamma$ point is observed in the underdoped cuprates at low temperatures\cite{gt3}. In YBCO and in Hg1201, Hall coefficient also changes sign from positive to negative at low temperatures\cite{gt3}, signifying the presence of an electron pocket. The origin of this electron pocket has been quite a debated question. In one view, the appearance of electron pocket at low temperatures is due to Fermi Surface Reconstruction (FSR) mechanism which originates in various charge ordering tendencies in the underdoped cuprates at low temperatures. However, Gor'kov and Teitel'baum (GT) differ on this. In their view it is permanent feature of underdoped cuprates and it is a band structure effect. 
 
 The question is why then the electron pocket becomes dominant only at low temperatures (sign change of the Hall coefficient)? In GT's view the main player is the mobility. Mobility of holes decrease as temperature is reduced in the underdoped regime. At a sufficiently low temperature, mobility of holes become less than that of electrons, and Hall coefficient changes its sign, and electron pocket appears\cite{gt3}.

 In GT's view there are two components (in the sense of electrons versus holes): (1) holes "resides" on Fermi "arcs" -- features along the nodal directions, and (2) electrons form an electron pocket at $\Gamma$ point.  {\it And temperature dependence of mobilities of these carriers is the main cause.}  In support of "electron pocket a band structure effect" they argue that there is residual metallic specific heat contribution deep inside the superconducting state of YBCO\cite{gt4,gt5}. However, in this author's opinion more quantitative analysis (taking into account the temperature dependence of mobilities of carriers) is needed to settle this issue. The concrete question to address is: why does the hole mobility decreases as temperature is lowered? To understand this one must investigate various mechanisms via which doped holes are scattered in underdoped cuprates.

\section{Some approaches that are not within the leeway of equation (1)}
 
Discussion of the approaches that fall outside the leeway of equation (1), and that attempt to address the Hall data in cuprates, is out of the scope of this mini review. For the purpose of completeness, we only mention them here so that an interested reader can refer to them. Referencing is in no way complete and author apologizes in advance (literature is simply huge).  

In 1996, Stojkovi\'{c} and Pines\cite{sp}, using the Nearly Antiferromagnetic Fermi Liquid (NAFL) description for planner quasiparticles with highly anisotropic scattering rate over the Fermi surface solved the Boltzmann equation (numerically). Their model could account for some of the features of the Hall data in optimally doped cuprates. The different temperature behaviour of Hall angle and $\rho_{ab}$ is connected with their choice of highly anisotropic scattering rate which scales as $\sim\frac{1}{T}$ over the "hot spots" and $\sim \frac{1}{T^2}$ over the corners of the Fermi surface. Ong and Anderson presented a critique of their approach\cite{oa}, to which Stojkovi\'{c} and Pines also responded\cite{sp2}. Full solution of the problem remained unsettled\cite{oa}. Kazuki Kanki and Hiroshi Kontani\cite{kh} also attempted to address these issues by taking into account current vertex corrections in addition to the introduction of the anisotropic scattering rates.  In 2003, Elihu Abrahams and C. M. Varma\cite{av} attempted to rationalize these issues using an extension of their Marginal Fermi Liquid Theory (MFLT) in which they incorporated a self-energy having both isotropic inelastic part and an anisotropic temperature independent elastic part. 

However, the physical and mathematical simplicity of the Gor'kov-teitel'baum (GT) approach is to be contrasted with the above approaches. GT's approach is able to bridge the gap between ARPES data and Hall data. The single most crucial aspect -- temperature dependence of the carrier density -- is taken into account in the GT approach. If the temperature dependence of the carrier density is a fundamental feature of the underdoped and optimally doped cuprates, then a large number of investigations of the anomalous properties of underdoped and optimally doped cuprates (which do not take this feature into account or address it as an outcome of the theory) may be on a shaky foundation. As suggested by GT, an attempt is made to understand Hall angle behaviour by taking into account the important effects of PG and resulting temperature dependence of carrier number density(\cite{nav3} and references therein).

The next step would be to build a microscopic foundation to this very simple but powerful phenomenological model. It has already been attempted. In a pioneering work\cite{shi}, Shiladitya Chakraborty and Philip Phillips (SP) build a microscopic foundation of this phenomenological model by starting from the Hubbard model. They start with the strong coupling limit ($U>>t$, where $U$ is Hubbard onsite interaction and $t$ is the tunneling matrix element in the Hubbard model), and a low energy effective action is obtained by integrating out the high energy physics (set by $U$) from where the Green's function and spectral function are computed. For the computation of $R_H = \frac{\sigma_{xy}}{\sigma_{xx}^2}$, the required longitudinal and transverse conductivities are computed using use the Kubo theory. From $R_H=\frac{1}{n_{Hall}(T) e}$ the carrier density $n_{Hall}(T)$ is obtained. They observe two-component behaviour thus providing a microscopic foundation to the GT phenomenological theory. However, an experimental detection of the charge-2e Boson (that is an inevitable consequence of the SP theory) or its effects remains an open issue.

\section{A comment on T-linear resistivity}

Resistivity of optimally doped cuprates exhibit linear-in-temperature behaviour over an extended temperature regime. It has been argued that it is due to the Planckian scattering rate which scales as $\frac{1}{\tau} \propto T$. Work of Gor'kov and Teitel'baum shows that within the leeway of equation (1) (which receives a direct verification in the underdoped LSCO as mentioned before), the temperature dependence of the carrier density cannot be neglected. Therefore, within the leeway of the Drude formula $\rho(T) = \frac{m^*}{n(T)e^2}\frac{1}{\tau(T)}$, T-linear resistivity does not imply T-linear scattering rate!  Approaches that are more fundamental (that goes beyond equation (1) and Drude formula of resistivity, such as the attempt by Stojkovi\'{c} and Pines\cite{sp2}) must address the temperature dependence of the Hall coefficient and T-linear resistivity, simultaneously (within the same setting), and must reduce to the phenomenological GTTA theory in the underdoped regime, at the minimum. This means that such an approach must take into account the effects of PG and temperature dependence of the carrier density. As far as author's knowledge goes this remains a challenging open problem.

\section{Summary}
In this short review, the main ideas of Gor'kov and Teitel'baum are presented. The Gor'kov-Teitel'baum Thermal Activation (GTTA) model is a successful model that captures the essential features of the PG state, including Hall data and PG phase boundary (when temperature independent term balances the temperature dependent thermal activation term in the GTTA model). The most beautiful thing regarding this model is its simplicity. And it bridges the gap between two completely different experiments (ARPES and Hall effect). The next step would be to investigate whether this model has a general validity and use beyond the scope of cuprates\cite{jal}.

\vspace{5cm}

\begin{acknowledgments}
Author would like to thank Philip Phillips for suggesting an important reference. He is also thankful to Mohit Randeria for extensive and illuminating discussions on various aspects of cuprates. 
\end{acknowledgments}

%---------------------------------------


\begin{thebibliography}{99}
\bibitem{fot1} For microscopic model, Anderson heuristically argued\cite{ander1} that the electronic state in cuprates is a 2D "Luttinger" liquid such that electron fractionalizes into "holons" and "spinons". Spinons are spin-half Fermions and spinon-spinon scattering leads to $T^2$ law like electron-electron scattering in a Fermi liquid.
\end{thebibliography}
\end{document}